# Sterile neutrinos existence suggested from LCT covariance

**Raoelina Andriambololona[1], Ravo Tokiniaina Ranaivoson[2], Hanitriarivo Rakotoson[3], Roland Raboanary[4]**

*raoelina.andriambololona@gmail.com[1], jacquelineraoelina@hotmail.com[1], raoelinasp@yahoofr[1], tokhiniaina@gmail.com[2], tokiniainaravor13@gmail.com[2], infotsara@gmail.com[3], r_raboanary@yahoo.fr[4]*

[1,2,3]*Information Technology and Theoretical Physics Department*
*Institut National des Sciences et Techniques Nucléaires ( INSTN- Madagascar)*
BP 3907 Antananarivo101, Madagascar, *instn@moov.mg*

[1, 2,3]TWAS *Madagascar Chapter, Malagasy Academy*
BP 4279 Antananarivo 101, Madagascar

[3,4]*Faculty of Sciences – University of Antananarivo*
BP 566 Antananarivo 101, Madagascar

**Corresponding author:** Ravo Tokiniaina Ranaivoson

*Abstract:* Sterile neutrinos are known to be hypothetical neutrinos which do not interact via the fundamental interactions described within the Standard Model of Particles Physics i.e. electroweak and strong interactions. They are expected to be important for the understanding of the physics beyond the current Standard Model. In the present work, it is shown that the existence of these particles can be suggested from covariance principle using a covariance group formed by Linear Canonical Transformations (LCTs) associated to a pentadimensional pseudo-Euclidian space. It is established that a spin representation of the LCT group gives a particle classification, applicable to the three families of leptons and quarks, which leads to the prediction of the existence of three sterile neutrinos and their antiparticles.

# 1-Introduction

Sterile neutrinos are currently considered to be very interesting because they are expected to be closely related to a new physics beyond the Standard Model and then to a unified theory of interactions including gravity. Some studies related to sterile neutrinos both on the theoretical and experimental aspects can be found for instance in the references [1-4].

In this work, our objective is to show that the existence of sterile neutrinos can be deduced from a covariance principle based on a covariance group formed by Linear Canonical Transformations (LCTs). In fact, it is established that a classification of the three families of leptons and quarks which predicts the existence of three sterile neutrinos (and their antiparticles) can be deduced from a spin representation of the LCT group corresponding to a pentadimensional pseudo-Euclidian space.

The mathematical framework considered all through this paper, is based on the formalism which was firstly developed in the reference [5]. Some of the main results established in this reference are then reviewed. The notation system used is taken from the reference [6]. Boldfaced letter like $\boldsymbol{p}$ corresponds to quantum operator $\boldsymbol{p}$ and normal letter $p$ corresponds to its eigenvalue. The natural unit system (the reduced Planck constant $\hbar = 1$ and the speed of light $c = 1$) is used too.

## 2- LCT group and its spin representation

LCTs are studied in optics, signal processing, and quantum mechanics [7-9]. As in the reference [5], we may define, in the framework of a relativistic quantum theory, the LCT corresponding to an $N$-dimensional pseudo-Euclidian space by the relations

$$\begin{cases} \boldsymbol{p}'_\mu = \mathbb{a}_\mu^\nu \boldsymbol{p}_\nu + \mathbb{b}_\mu^\nu \boldsymbol{x}_\nu \\ \boldsymbol{x}'_\mu = \mathbb{c}_\mu^\nu \boldsymbol{p}_\nu + \mathbb{d}_\mu^\nu \boldsymbol{x}_\nu \\ [\boldsymbol{p}'_\mu, \boldsymbol{x}'_\nu] = [\boldsymbol{p}_\mu, \boldsymbol{x}_\nu] = i\eta_{\mu\nu} \\ [\boldsymbol{p}'_\mu, \boldsymbol{p}'_\nu] = [\boldsymbol{p}_\mu, \boldsymbol{p}_\nu] = 0 \\ [\boldsymbol{x}'_\mu, \boldsymbol{x}'_\nu] = [\boldsymbol{x}_\mu, \boldsymbol{x}_\nu] = 0 \end{cases} \quad (1)$$

in which $\boldsymbol{p}_\mu$ and $\boldsymbol{x}_\mu$ are the momenta and coordinates operators associated to a first frame of reference while $\boldsymbol{p}'_\mu$ and $\boldsymbol{x}'_\mu$ are their analogs associated to a second frame of reference. The two frames of reference are linked by the LCT. The $\eta_{\mu\nu}$ are the covariant components of the bilinear form corresponding to the scalar product on the considered pseudo-Euclidian space.

The relations in (1) mean that LCTs can be identified to be the linear transformations which leave invariant the Canonical Commutation Relations which characterize the coordinates and momenta operators.

The relations in (1) imply that the $N \times N$ matrices $\mathbb{a}, \mathbb{b}, \mathbb{c}, \mathbb{d}$ corresponding to the parameters $\mathbb{a}_\mu^\nu, \mathbb{b}_\mu^\nu, \mathbb{c}_\mu^\nu, \mathbb{d}_\mu^\nu$ fulfill the following relations

$$\begin{cases} \mathbb{a}^T \eta \mathbb{d} - \mathbb{b}^T \eta \mathbb{c} = \eta \\ \mathbb{a}^T \eta \mathbb{b} - \mathbb{b}^T \eta \mathbb{a} = 0 \\ \mathbb{c}^T \eta \mathbb{d} - \mathbb{d}^T \eta \mathbb{c} = 0 \end{cases} \Leftrightarrow \begin{pmatrix} \mathbb{a} & \mathbb{c} \\ \mathbb{b} & \mathbb{d} \end{pmatrix}^T \begin{pmatrix} 0 & \eta \\ -\eta & 0 \end{pmatrix} \begin{pmatrix} \mathbb{a} & \mathbb{c} \\ \mathbb{b} & \mathbb{d} \end{pmatrix} = \begin{pmatrix} 0 & \eta \\ -\eta & 0 \end{pmatrix} \quad (2)$$

The relations in (2) mean that the set of the $2N \times 2N$ matrices $\mathbb{t} = \begin{pmatrix} \mathbb{a} & \mathbb{c} \\ \mathbb{b} & \mathbb{d} \end{pmatrix}$ can be identified with the symplectic group $Sp(2N_+, 2N_-)$ [5, 10]. Let us denote this group by $\mathbb{T}$.

$$\mathbb{T} = \{\mathfrak{t} = \begin{pmatrix} \mathfrak{a} & \mathfrak{c} \\ \mathfrak{b} & \mathfrak{d} \end{pmatrix} / \begin{pmatrix} \mathfrak{a} & \mathfrak{c} \\ \mathfrak{b} & \mathfrak{d} \end{pmatrix}^T \begin{pmatrix} 0 & \eta \\ -\eta & 0 \end{pmatrix} \begin{pmatrix} \mathfrak{a} & \mathfrak{c} \\ \mathfrak{b} & \mathfrak{d} \end{pmatrix} = \begin{pmatrix} 0 & \eta \\ -\eta & 0 \end{pmatrix}\} \cong Sp(2N_+, 2N_-) \quad (3)$$

If an LCT group $\mathbb{T}$ is taken to be a covariance group for a relativistic quantum theory, it is expected that the law of transformations of the quantities that are considered, in the framework of this theory, should correspond to some representations of $\mathbb{T}$. It can be shown that the fermionic fields can be considered as corresponding to a spinorial representation.

In [5], it was established that a surjective group homomorphism $f$ can be defined between the LCT group $\mathbb{T}$ and a group $\mathbb{G}$ corresponding to the following group isomorphism

$$\mathbb{G} \cong Sp(2N_+, 2N_-) \cap SO_0(2N_+, 2N_-) \cong U(N_+, N_-) \quad (4)$$

According to this relation (4), $\mathbb{G}$ is a subgroup of the identinty component $SO_0(2N_+, 2N_-)$ of the special orthogonal group $SO(2N_+, 2N_-)$. It follows that $f$ defines a special orthogonal representation of LCTs.

A spinorial representation of the LCT group $\mathbb{T}$ can be deduced by taking advantage of the relation between $SO_0(2N_+, 2N_-)$ and its topological double cover i.e. the spin group $Spin(2N_+, 2N_-)$ [11-12]. The relation between $\mathbb{G}$ and its double cover, denoted $\mathbb{S}$, which is a subgroup of $Spin(2N_+, 2N_-)$, is described by a covering map $u$. $u$ is surjective: an element $\mathcal{S}$ of $\mathbb{S}$ and its opposite $-\mathcal{S}$ have the same image $\mathfrak{g}$.

$$\begin{cases} u: \mathbb{S} \to \mathbb{G} \\ \mathcal{S} \mapsto \mathfrak{g} \end{cases} \quad Ker(u) \cong \{1, -1\} \quad (5)$$

Given the existence of the group homomorphism $f$ between $\mathbb{G}$ and $\mathbb{T}$, $u$ defines a spinorial representation of LCTs.

Some of the main properties of $\mathbb{G} \cong U(N_+, N_-)$ are given in the existing literature related to the pseudo-unitary group $U(N_+, N_-)$ [13-14]. Some of the main properties of the double cover $\mathbb{S}$ of $\mathbb{G}$ follows from these properties of $\mathbb{G}$. It can be, for instance, founded that for the signature $(N_+, N_-) = (1,4)$, $\mathbb{G}$ and $\mathbb{S}$ are connected and their dimension is equal to 25. The dimension of their maximal compact subgroups is equal to 17. The rank of these maximal compact subgroups is equal to 5.

As shown in [5], a basis of the Lie algebra $\mathfrak{s}$ of the Lie group $\mathbb{S}$ is the family

$$\mathfrak{B} = \{\frac{1}{2}(\alpha^\mu \alpha^\nu + \beta^\mu \beta^\nu), \frac{1}{2}(\alpha^\mu \beta^\nu + \alpha^\nu \beta^\mu)\} \quad \mu = 0,1,2,3,4 \quad (6)$$

in which $\alpha^\mu$ and $\beta^\mu$ are the generators of the Clifford algebra $\mathcal{C}\ell(2,8)$. For the algebra corresponding to the compact part of $\mathbb{S}$, the basis is

$$\mathfrak{B}_c = \{\frac{1}{2}(\alpha^j \alpha^l + \beta^j \beta^l), \alpha^\mu \beta^\mu, \frac{1}{2}(\alpha^j \beta^l + \alpha^l \beta^j)\} \quad j,l = 1,2,3,4 \quad (7)$$

**3-Classification of leptons and quarks and identification of the sterile neutrinos**

The dimension of the Cartan subalgebra of the algebra generated by the set $\mathfrak{B}_c$ given in (7) is equal to 5. The following five commutative hermitian generators can be chosen to be a basis of this Cartan subalgebra

$$\mathcal{Y}^0 = \frac{1}{2}i\alpha^0\beta^0 \quad \mathcal{Y}^1 = \frac{1}{3}i\alpha^1\beta^1 \quad \mathcal{Y}^2 = \frac{1}{3}i\alpha^2\beta^2 \quad \mathcal{Y}^3 = \frac{1}{3}i\alpha^3\beta^3 \quad \mathcal{Y}^4 = \frac{1}{2}i\alpha^4\beta^4 \quad (8)$$

An explicit matrices representations of these operators can be obtained if an explicit matrices representation of the operators $\alpha^\mu$ and $\beta^\mu$ are given. A choice that can be considered for our purpose is ($\sigma^0$ is the $2 \times 2$ identity matrix and $\sigma^1, \sigma^2, \sigma^3$ are the well-known Pauli matrices)

$$\begin{cases} \alpha^0 = \sigma^1 \otimes \sigma^0 \otimes \sigma^0 \otimes \sigma^0 \otimes \sigma^0 & \beta^0 = \sigma^2 \otimes \sigma^0 \otimes \sigma^0 \otimes \sigma^0 \otimes \sigma^0 \\ \alpha^1 = i\sigma^3 \otimes \sigma^1 \otimes \sigma^0 \otimes \sigma^0 \otimes \sigma^0 & \beta^1 = -i\sigma^3 \otimes \sigma^2 \otimes \sigma^0 \otimes \sigma^0 \otimes \sigma^0 \\ \alpha^2 = i\sigma^3 \otimes \sigma^3 \otimes \sigma^1 \otimes \sigma^0 \otimes \sigma^0 & \beta^2 = -i\sigma^3 \otimes \sigma^3 \otimes \sigma^2 \otimes \sigma^0 \otimes \sigma^0 \\ \alpha^3 = i\sigma^3 \otimes \sigma^3 \otimes \sigma^3 \otimes \sigma^1 \otimes \sigma^0 & \beta^3 = -i\sigma^3 \otimes \sigma^3 \otimes \sigma^3 \otimes \sigma^2 \otimes \sigma^0 \\ \alpha^4 = i\sigma^3 \otimes \sigma^3 \otimes \sigma^3 \otimes \sigma^3 \otimes \sigma^1 & \beta^4 = -i\sigma^3 \otimes \sigma^3 \otimes \sigma^3 \otimes \sigma^3 \otimes \sigma^2 \end{cases} \quad (9)$$

To deduce the classification of leptons and quarks, the operators corresponding to the particles properties, as defined within the Standard model of particle physics should be introduced. These properties are the weak isospin, the weak hypercharge, the electric charge and strong color. According to the table 1 (see next page), an adequate choice for the operators $I_3, Y_W, Q$ corresponding respectively to the weak isospin, weak hypercharges and electric charges are

$$\begin{cases} I_3 = \frac{1}{2}\mathcal{Y}^0 - \frac{1}{2}\mathcal{Y}^4 \quad Y_W = \mathcal{Y}^0 + \mathcal{Y}^1 + \mathcal{Y}^2 + \mathcal{Y}^3 + \mathcal{Y}^4 \\ \quad Q = \mathcal{Y}^0 + \frac{1}{2}\mathcal{Y}^1 + \frac{1}{2}\mathcal{Y}^2 + \frac{1}{2}\mathcal{Y}^3 = I_3 + \frac{Y_W}{2} \end{cases} \quad (10)$$

The Table 1 shows also that the existence of strong colors can be explained through the possible combinations of the eigenvalues of the operators $\mathcal{Y}^1, \mathcal{Y}^2$ and $\mathcal{Y}^3$.

The Table 1 gives a classification of the three families of leptons and quarks according to the values of the eigenvalues of the operators $\mathcal{Y}^0, \mathcal{Y}^1, \mathcal{Y}^2, \mathcal{Y}^3, \mathcal{Y}^4, I_3, Y_W$ and $Q$. The up, charm and top quarks are denoted $u, c, t$ and the down, strange and bottom are denoted $d, s$ and $b$. Their antiparticles are respectively denoted $\bar{u}, \bar{c}, \bar{t}, \bar{d}, \bar{s}$ and $\bar{b}$. The lower index indicates the chirality: $R$ for right-handed and $L$ for left-handed. The upper index indicates the strong colors (blue, green or red). The charged lepton: electron, muon and tau are denoted $e, \mu$ and $\tau$ and their antiparticles are denoted $\bar{e}, \bar{\mu}$ and $\bar{\tau}$. The corresponding neutrinos are respectively denoted $\nu_e, \nu_\mu$ and $\nu_\tau$ and the antineutrinos are denoted $\bar{\nu}_e, \bar{\nu}_\mu$ and $\bar{\nu}_\tau$.

The Table 1 suggests the existence of the three sterile neutrinos $\nu_{eR}, \nu_{\mu R}, \nu_{\tau R}$ and their antiparticles $\bar{\nu}_{eR}, \bar{\nu}_{\mu R}$ and $\bar{\nu}_{\tau R}$ (put in boldfaced letters). Their electric charge, isospin and hypercharge are equal to zero as expected.

The Table 1 suggests also the possibility of description of a family of fermions with a single field $\psi$ [5]. The law of transformations of this field under the action of an LCT is described by the element $\mathcal{S}$ of the group $\mathbb{S}$ defined through the spinorial representation (5)

$$\psi' = \mathcal{S}\psi \quad (11)$$

The relation (11) shows explicitly, as expected, that a fermionic field $\psi$ transforms covariantly under the action of the LCTs corresponding to $\mathcal{S}$. The law of transformation is described through the spin representation of the LCT group $\mathbb{T}$ which is the covariance group.

Table 1: *Classification of quarks, leptons and their antiparticles with the sterile neutrinos.*

| N° | \mathcal{Y}^0 | \mathcal{Y}^1 | \mathcal{Y}^2 | \mathcal{Y}^3 | \mathcal{Y}^4 | $I_3$ | $Y_W$ | $Q$ | 1st | 2nd | 3rd |
|---|---|---|---|---|---|---|---|---|---|---|---|
| | | | | OPERATORS EIGENVALUES | | | | | PARTICLES FAMILY | | |
| 1 | -1/2 | -1/3 | -1/3 | -1/3 | -1/2 | 0 | -2 | -1 | $e_R$ | $\mu_R$ | $\tau_R$ |
| 2 | -1/2 | -1/3 | -1/3 | -1/3 | 1/2 | -1/2 | -1 | -1 | $e_L$ | $\mu_L$ | $\tau_L$ |
| 3 | -1/2 | -1/3 | -1/3 | 1/3 | -1/2 | 0 | -4/3 | -2/3 | $\bar{u}_R^{blue}$ | $\bar{c}_R^{blue}$ | $\bar{t}_R^{blue}$ |
| 4 | -1/2 | -1/3 | -1/3 | 1/3 | 1/2 | -1/2 | -1/3 | -2/3 | $\bar{u}_L^{blue}$ | $\bar{c}_L^{blue}$ | $\bar{t}_L^{blue}$ |
| 5 | -1/2 | -1/3 | 1/3 | -1/3 | -1/2 | 0 | -4/3 | -2/3 | $\bar{u}_R^{green}$ | $\bar{c}_R^{green}$ | $\bar{t}_R^{green}$ |
| 6 | -1/2 | -1/3 | 1/3 | -1/3 | 1/2 | -1/2 | -1/3 | -2/3 | $\bar{u}_L^{green}$ | $\bar{c}_L^{green}$ | $\bar{t}_L^{green}$ |
| 7 | -1/2 | -1/3 | 1/3 | 1/3 | -1/2 | 0 | -2/3 | -1/3 | $d_R^{red}$ | $s_R^{red}$ | $b_R^{red}$ |
| 8 | -1/2 | -1/3 | 1/3 | 1/3 | 1/2 | -1/2 | 1/3 | -1/3 | $d_L^{red}$ | $s_L^{red}$ | $b_L^{red}$ |
| 9 | -1/2 | 1/3 | -1/3 | -1/3 | -1/2 | 0 | -4/3 | -2/3 | $\bar{u}_R^{red}$ | $\bar{c}_R^{red}$ | $\bar{t}_R^{red}$ |
| 10 | -1/2 | 1/3 | -1/3 | -1/3 | 1/2 | -1/2 | -1/3 | -2/3 | $\bar{u}_L^{red}$ | $\bar{c}_L^{red}$ | $\bar{t}_L^{red}$ |
| 11 | -1/2 | 1/3 | -1/3 | 1/3 | -1/2 | 0 | -2/3 | -1/3 | $d_R^{green}$ | $s_R^{green}$ | $b_R^{green}$ |
| 12 | -1/2 | 1/3 | -1/3 | 1/3 | 1/2 | -1/2 | 1/3 | -1/3 | $d_L^{green}$ | $s_L^{green}$ | $b_L^{green}$ |
| 13 | -1/2 | 1/3 | 1/3 | -1/3 | -1/2 | 0 | -2/3 | -1/3 | $d_R^{blue}$ | $s_R^{blue}$ | $b_R^{blue}$ |
| 14 | -1/2 | 1/3 | 1/3 | -1/3 | 1/2 | -1/2 | 1/3 | -1/3 | $d_L^{blue}$ | $s_L^{blue}$ | $b_L^{blue}$ |
| **15** | **-1/2** | **1/3** | **1/3** | **1/3** | **-1/2** | **0** | **0** | **0** | $\bar{\nu}_{eR}$ | $\bar{\nu}_{\mu R}$ | $\bar{\nu}_{\tau R}$ |
| 16 | -1/2 | 1/3 | 1/3 | 1/3 | 1/2 | -1/2 | 1 | 0 | $\bar{\nu}_{eL}$ | $\bar{\nu}_{\mu L}$ | $\bar{\nu}_{\tau L}$ |
| 17 | 1/2 | -1/3 | -1/3 | -1/3 | -1/2 | 1/2 | -1 | 0 | $\nu_{eL}$ | $\nu_{\mu L}$ | $\nu_{\tau L}$ |
| **18** | **1/2** | **-1/3** | **-1/3** | **-1/3** | **1/2** | **0** | **0** | **0** | $\nu_{eR}$ | $\nu_{\mu R}$ | $\nu_{\tau R}$ |
| 19 | 1/2 | -1/3 | -1/3 | 1/3 | -1/2 | 1/2 | -1/3 | 1/3 | $\bar{d}_L^{blue}$ | $\bar{s}_L^{blue}$ | $\bar{b}_L^{blue}$ |
| 20 | 1/2 | -1/3 | -1/3 | 1/3 | 1/2 | 0 | 2/3 | 1/3 | $\bar{d}_R^{blue}$ | $\bar{s}_R^{blue}$ | $\bar{b}_R^{blue}$ |
| 21 | 1/2 | -1/3 | 1/3 | -1/3 | -1/2 | 1/2 | -1/3 | 1/3 | $\bar{d}_L^{gree}$ | $\bar{s}_L^{gree}$ | $\bar{b}_L^{gree}$ |
| 22 | 1/2 | -1/3 | 1/3 | -1/3 | 1/2 | 0 | 2/3 | 1/3 | $\bar{d}_R^{green}$ | $\bar{s}_R^{green}$ | $\bar{b}_R^{green}$ |
| 23 | 1/2 | -1/3 | 1/3 | 1/3 | -1/2 | 1/2 | 1/3 | 2/3 | $u_L^{red}$ | $c_L^{red}$ | $t_L^{red}$ |
| 24 | 1/2 | -1/3 | 1/3 | 1/3 | 1/2 | 0 | -4/3 | 2/3 | $u_R^{red}$ | $c_R^{red}$ | $t_R^{red}$ |
| 25 | 1/2 | 1/3 | -1/3 | -1/3 | -1/2 | 1/2 | 1/3 | 1/3 | $\bar{d}_L^{red}$ | $\bar{s}_L^{red}$ | $\bar{b}_L^{red}$ |
| 26 | 1/2 | 1/3 | -1/3 | -1/3 | 1/2 | 0 | 2/3 | 1/3 | $\bar{d}_R^{red}$ | $\bar{s}_R^{red}$ | $\bar{b}_R^{red}$ |
| 27 | 1/2 | 1/3 | -1/3 | 1/3 | -1/2 | 1/2 | 1/3 | 2/3 | $u_L^{green}$ | $c_L^{green}$ | $t_L^{green}$ |
| 28 | 1/2 | 1/3 | -1/3 | 1/3 | 1/2 | 0 | -4/3 | 2/3 | $u_R^{green}$ | $c_R^{green}$ | $t_R^{green}$ |
| 29 | 1/2 | 1/3 | 1/3 | -1/3 | -1/2 | 1/2 | 1/3 | 2/3 | $u_L^{blue}$ | $c_L^{blue}$ | $t_L^{blue}$ |
| 30 | 1/2 | 1/3 | 1/3 | -1/3 | 1/2 | 0 | -4/3 | 2/3 | $u_R^{blue}$ | $c_R^{blue}$ | $t_R^{blue}$ |
| 31 | 1/2 | 1/3 | 1/3 | 1/3 | -1/2 | 1/2 | 1 | 1 | $\bar{e}_L$ | $\bar{\mu}_L$ | $\bar{\tau}_L$ |
| 32 | 1/2 | 1/3 | 1/3 | 1/3 | 1/2 | 0 | 2 | 1 | $\bar{e}_R$ | $\bar{\mu}_R$ | $\bar{\tau}_R$ |

## 4- Discussion and Conclusion

It is highlighted through this work that a spinorial representation of an LCT group, considered as covariance group, can lead to a description of the charges of the elementary fermions of the Standard Model and to the prediction of the existence of three sterile neutrinos and their antiparticles. The LCTs to be considered for this purpose are those which correspond to a pentadimensional pseudo-Euclidian space.

The LCT group may be also used to be a gauge group for a unified theory of interaction including gravity. It is, in fact, remarked in [5] that the symmetry associated to LCT covariance, which is a quantum phase space symmetry, can be considered to circumvent the Coleman-Mandula no-go theorem [15-16]. And the main gauge groups that are currently considered in the gauge theories of gravitation i.e. the Poincaré and de Sitter groups [17-18] can be obtained from the contraction of an LCT group [5]. This contraction is to be understood within the concept of group contraction introduced by Inönü and Wigner [19-23].

These results and facts show that the topics of sterile neutrinos and unified theory of interactions (which include gravity) are closely related as it may be expected and this relation seems to be exactly described through LCT covariance. Further studies are needed to test or confirm this point view. An in-depth study of the exact relation between the LCT group and the gauge group of the Standard Model may be for instance firstly considered to analyze, among other things, the case of the bosonic sector. Indeed, the study which is considered in this paper was oriented on fermions.

**Acknowledgements:** The authors are grateful to the INSTN-Madagascar and its staff for their help during the conception and completion of this work and to the reviewers for their interesting comments and suggestions.